\documentclass[11pt]{article}
\usepackage{amssymb}

\newcommand{\boldsymbol}{\bf}
\newcommand{\BE}{\begin{equation}}
\newcommand{\EE}{\end{equation}}

\begin{document}

%\begin{titlepage}

\title{Is the physical vacuum a preferred frame ?}

\author{M. Consoli  and E. Costanzo\\
Istituto Nazionale di Fisica Nucleare, Sezione di Catania \\
Dipartimento di Fisica e Astronomia dell' Universit\`a di Catania}

\maketitle

\begin{abstract}
It is generally assumed that the physical vacuum of particle physics
should be characterized by an energy momentum tensor in such a way
to preserve exact Lorentz invariance. On the other hand, if the
ground state were characterized by its energy-momentum vector, with
zero spatial momentum and a non-zero energy, the vacuum would
represent a preferred frame. Since both theoretical approaches have
their own good motivations, we
propose an experimental test to decide between the two scenarios.  
\end{abstract}

\vskip 20pt

\section{Introduction}

The possible existence of a preferred reference frame $\Sigma$ is an
old and important issue that dates back to the origin of the theory
of relativity and to the basic differences between Einstein's
special relativity and the Lorentzian point of view. No doubt,
today, the former interpretation is widely accepted. However, in
spite of the deep conceptual differences, it is not so obvious how
to distinguish experimentally the two interpretations.

For a modern presentation of the Lorentzian approach one can, for
instance, follow Bell \cite{bell}. Differently from the usual
derivations within special relativity, one starts from physical
modifications of matter (namely Larmor's time dilation and
Lorentz-Fitzgerald length contraction) to deduce the basic Lorentz
transformation between $\Sigma$ and any moving frame S'. Due to the
crucial underlying group property, two observers S' and S'',
individually connected to $\Sigma$ by a Lorentz transformation, are
then also mutually connected by a Lorentz transformation with
relative velocity parameter fixed by the velocity composition rule.
As a consequence, one deduces a substantial quantitative equivalence
of the two formulations of relativity for most standard experimental
tests.

Thus, one is naturally driven back to the old question: if there
were a preferred frame $\Sigma$, could one observe the motion with
respect to it ? In Sect.2, after reviewing the general problem of
vacuum condensation in present particle physics, we shall argue that
this might indeed be possible. In fact, by accepting the idea that
the physical vacuum might be defined by its energy-momentum vector,
with zero spatial momentum and non-zero energy, one deduces that
such a vacuum represent a preferred frame since any moving observer
should feel an energy-momentum flow along the direction of motion.

After this first part, we shall consider in Sect.3 an alternative
point of view where the vacuum is only characterized by a suitable
expectation value of the energy-momentum tensor and the previous
conclusion is not true.

Since the two theoretical approaches have their own good
motivations, we shall compare phenomenologically the two scenarios.
This other part will be discussed in Sects.4 and 5 where a possible
experimental test will be proposed. Finally, Sect.6 will contain a
summary and our conclusions.

\vskip 10 pt

\section{Vacuum energy and Lorentz invariance}

The phenomenon of vacuum condensation, that has changed
substantially the old view of the vacuum in axiomatic quantum field
theory \cite{cpt}, in the physically relevant case of the Standard
Model of electroweak interactions can be summarized saying that
``What we experience as empty space is nothing but the configuration
of the Higgs field that has the lowest possible energy. If we move
from field jargon to particle jargon, this means that empty space is
actually filled with Higgs particles. They have Bose condensed"
\cite{thooft}. In the simplified case of a pure $\Phi^4$ theory,
this condensation phenomenon can be explicitly checked by
constructing \cite{cian} a variational approximation to the
spontaneously broken phase as a coherent state built up with the
creation and annihilation operators of an empty reference vacuum
state $|0\rangle$. Thus, it becomes natural to ask \cite{pagano} if
the macroscopic occupation of the same quantum state, i.e.
${\boldsymbol{k}}=0$ in some reference frame $\Sigma$, could
represent the operative construction of a ``quantum ether". This
would characterize the {\it physically realized} form of relativity
and could play the role of preferred frame in a modern Lorentzian
approach.

Usually this possibility is not considered with the motivation,
perhaps, that the average properties of the condensed phase are
summarized into a single quantity that transforms as a world scalar
under the Lorentz group (in the Standard Model, the vacuum
expectation value $\langle\Phi\rangle$ of the Higgs field). However,
this does not imply that the physical vacuum state has to be Lorentz
invariant. Namely, Lorentz transformation operators ${U}'$,
${U}''$,..might transform non trivially the basic vacuum state
$|\Psi^{(0)}\rangle$ (appropriate to an observer at rest in
$\Sigma$) into new vacuum states $| \Psi'\rangle$, $|
\Psi''\rangle$,.. (appropriate to moving observers S', S'',..) and
still, for any Lorentz-invariant operator ${G}$, one would find
\begin{equation} \langle {G}\rangle_{\Psi^{(0)}}=\langle
{G}\rangle_{\Psi'}=\langle {G}\rangle_{\Psi''}=..\end{equation} As a
matter of fact, this view of a non-Lorentz-invariant vacuum turns
out to be unavoidable when combining the general idea of a non-zero
vacuum energy with the algebra of the 10 generators $P_\alpha$,
$M_{\alpha,\beta}$ ( $\alpha$ ,$\beta$=0, 1, 2, 3) of the Poincar\'e
group. Here $P_\alpha$ are the 4 generators of the space-time
translations and $M_{\alpha\beta}=-M_{\beta\alpha}$ are the  6
generators of the Lorentzian rotations with the following
commutation relations \begin{equation} \label{tras1}
[P_\alpha,P_\beta]=0 \end{equation} \begin{equation} \label{boost}
[M_{\alpha\beta}, P_\gamma]= \eta_{\beta\gamma}P_\alpha -
\eta_{\alpha\gamma}P_\beta \end{equation} \begin{equation}
\label{tras2} [M_{\alpha\beta}, M_{\gamma\delta}]=
\eta_{\alpha\gamma}M_{\beta\delta}+
\eta_{\beta\delta}M_{\alpha\gamma}
-\eta_{\beta\gamma}M_{\alpha\delta}-\eta_{\alpha\delta}M_{\beta\gamma}
\end{equation} where $\eta_{\alpha\beta}={\rm diag}(1,-1,-1,-1)$.

In the following we shall assume the existence of a suitable
operatorial representation of the Poincar\'e algebra for the quantum
theory (where $\hat{M}_{32}=-i\hat{J}_1$, $\hat{M}_{01}=-i\hat{L}_1$
and cyclic permutations). We shall also assume that the occurrence
of vacuum condensation does not modify the structure of the basic
commutation relations (\ref{tras1})$-$(\ref{tras2}). This
hypothesis, that reflects the observed properties of the
energy-momentum under Lorentz transformations and as such has a
sound experimental basis, is also completely consistent with the
general attitude toward the phenomenon of spontaneous symmetry
breaking. This, while giving rise to a non-symmetric ground state,
leaves unchanged the commutation relations among the generators of
the underlying dynamical symmetry group.

Within this framework, one possible assumption behind a non-trivial
vacuum is that the physical vacuum state $|\Psi^{(0)}\rangle$
\cite{degenerate} maintains both zero momentum and zero angular
momentum, i.e. (i,j=1,2,3)\begin{equation} \label{groundi}
\hat{P}_i|\Psi^{(0)}\rangle=\hat{M}_{ij}|\Psi^{(0)}\rangle=0
\end{equation} but, at the same time, is characterized by a
non-vanishing energy
\begin{equation} \label{ground}
\hat{P}_0|\Psi^{(0)}\rangle=E_0|\Psi^{(0)}\rangle \end{equation}
This vacuum energy might have very different explanations. Here, we
shall limit ourselves to explore the physical implications of its
existence by just observing that, in interacting quantum field
theories, there is no known way to ensure consistently the condition
$E_0=0$ without imposing an unbroken supersymmetry (which is not
phenomenologically acceptable \cite{conformal}).

To this end, let us consider the generator of a
Lorentz-transformation along the 1-axis $\hat{M}_{01}$. For $E_0\neq
0$, finite Lorentz transformations obtained by exponentiating
$\hat{M}_{01}$ will produce new vacuum states $| \Psi'\rangle$, $|
\Psi''\rangle$,..that differ  non trivially from
$|\Psi^{(0)}\rangle$. They maintain zero eigenvalues of $\hat{P}_2$
and $\hat{P}_3$ (since these commute with $\hat{M}_{01}$) while the
mean value of $\hat{P}_1$ and $\hat{P}_0$ can be computed by
defining a boosted vacuum state $| \Psi'\rangle$ as
\begin{equation}\label{trasf1} | \Psi'\rangle=
e^{\lambda'\hat{M}_{01}}|\Psi^{(0)}\rangle \end{equation} (recall
that $\hat{M}_{01}=-i\hat{L}_1$ is an anti-hermitian operator) and
using the relations \begin{equation} \label{po1}
e^{-\lambda'\hat{M}_{01}}~\hat{P}_1~
e^{\lambda'\hat{M}_{01}}=\cosh\lambda'~\hat{P}_1 +
\sinh\lambda'~\hat{P}_0
\end{equation} \begin{equation} \label{po2} e^{-\lambda'\hat{M}_{01}}~\hat{P}_0
~e^{\lambda'\hat{M}_{01}}=\sinh\lambda'~\hat{P}_1 +
\cosh\lambda'~\hat{P}_0
\end{equation}  In this way, one finds \begin{equation}
\label{boost1} \langle
{\hat{P}_1}\rangle_{\Psi'}=E_0\sinh\lambda'\end{equation}
\begin{equation} \label{boost2} \langle
{\hat{P}_0}\rangle_{\Psi'}=E_0\cosh\lambda'\end{equation} so that a
boost produces a vacuum energy-momentum flow along the direction of
motion with respect to $\Sigma$. Therefore, in the spirit of both
classical and quantum field theory, where global quantities are
obtained by integrating local densities over 3-space, for a moving
observer S' the physical vacuum looks like some kind of ethereal
medium for which, in general, one can introduce a momentum density
$W_{0i}$ through the relation (i=1,2,3)
\begin{equation} \label{density} \langle {\hat{P}_i}\rangle_{\Psi'}=\int
d^3x~W_{0i} \neq 0 \end{equation}

\section{The energy-momentum tensor of the vacuum}

There is however another approach to the problem of the vacuum that
leads to completely different conclusions. According to
refs.\cite{zeldovich,weinberg}, the physical vacuum state
$|\Psi^{(0)}\rangle$ should not be considered an eigenstate of the
energy-momentum operator but should rather be characterized by the
expectation value of the local energy-momentum tensor
$\hat{W}_{\mu\nu}$. Since the only Lorentz-invariant tensor is
$\eta_{\mu\nu}$, this should have the form
\begin{equation} \label{zeld} \langle
\hat{W}_{\mu\nu}\rangle_ {\Psi^{(0)}}=\rho
~\eta_{\mu\nu}\end{equation} $\rho$ being a space-time independent
constant. In this case, by introducing the Lorentz transformation
matrices $\Lambda^\mu_\nu$ to any moving frame S', defining $\langle
\hat{W}_{\mu\nu}\rangle_{\Psi'}$ through the relation \BE\langle
\label{cov}
\hat{W}_{\mu\nu}\rangle_{\Psi'}=\Lambda^{\sigma}{_\mu}\Lambda^{\rho}{_\nu}
~\langle\hat{W}_{\sigma\rho}\rangle_{\Psi^{(0)}}\EE and using
eq.(\ref{zeld}),
 it follows that the expectation value of $\hat{W}_{0i}$ in any
 boosted vacuum state $| \Psi'\rangle$ vanishes, just as it vanishes
in $|\Psi^{(0)}\rangle$. Therefore, differently from
eq.(\ref{density}), one gets
\begin{equation} \label{density1} \langle
{\hat{P}_i}\rangle_{\Psi'}=\int d^3x~ \langle
\hat{W}_{0i}\rangle_{\Psi'}= 0 \end{equation} To resolve the
conflict, the author of ref.\cite{zeldovich} advocates the point of
view that the vacuum energy $E_0$ is likely infinite and represents
a spurious concept. Thus one should definitely replace
eqs.(\ref{groundi})-(\ref{ground}) with eq.(\ref{zeld}) (``the
question is not whether the vacuum has an energy-momentum vector but
whether the vacuum has an energy-momentum tensor").

The issue is non trivial and does not possess a simple solution. We
can only observe that, by accepting this point of view, one might be
faced with some consistency problems. For instance, in a
second-quantized formalism, single-particle energies $E_1({\bf{p}})$
are defined as the energies of the corresponding one-particle states
$|{\bf{p}}\rangle$ minus the energy of the zero-particle, vacuum
state. If $E_0$ is considered a spurious concept, also
$E_1({\bf{p}})$ will become an ill-defined quantity.

At a deeper level, one should also realize that in an approach based
only on eq.(\ref{zeld}) the properties of $|\Psi^{(0)}\rangle$ under
a Lorentz transformation are not well defined. In fact, a
transformed vacuum state $| \Psi'\rangle$ is obtained, for instance,
by acting on $|\Psi^{(0)}\rangle$ with the boost generator
$\hat{M}_{01}$ as in eq.(\ref{trasf1}). Once $|\Psi^{(0)}\rangle$ is
considered an eigenstate of the energy-momentum operator as in
Sect.2, one can definitely show that, for $E_0\neq 0$, $|
\Psi'\rangle$ and $|\Psi^{(0)}\rangle$ differ non-trivially. On the
other hand, if $E_0=0$ there are only two alternatives: either
$\hat{M}_{01}|\Psi^{(0)}\rangle=0$, so that
$|\Psi'\rangle=|\Psi^{(0)}\rangle$,  or
$\hat{M}_{01}|\Psi^{(0)}\rangle$ is a state vector proportional to
$|\Psi^{(0)}\rangle$, so that $| \Psi'\rangle$ and
$|\Psi^{(0)}\rangle$ differ by a phase factor.

Therefore, if the structure in eq.(\ref{zeld}) were really
equivalent to the exact Lorentz invariance of the vacuum, it should
be possible to show similar results, for instance that such a
$|\Psi^{(0)}\rangle$ state can remain invariant under a boost, i.e.
be an eigenstate of \BE \hat{M}_{0i}=-i\int
d^3x~(x_i\hat{W}_{00}-x_0 \hat{W}_{0i}) \EE with zero eigenvalue. As
far as we can see, there is no way to obtain such a result by just
starting from eq.(\ref{zeld}) (that only amounts to the weaker
condition $\langle \hat{M}_{0i}\rangle_ {\Psi^{(0)}}=0$). Thus,
independently of the finiteness of $E_0$, it should not come as a
surprise that one can run into contradictory statements once
$|\Psi^{(0)}\rangle$ is instead characterized by means of
eqs.(\ref{groundi})-(\ref{ground}).

For these reasons, it is not so obvious that the local relations
(\ref{zeld}) represent a more fundamental approach to the vacuum, as
compared to our previous analysis in Sect.2. Rather, in our opinion,
both approaches have their own good motivations and, to decide
between eqs.(\ref{density}) and (\ref{density1}), one should try to
work out the possible observable consequences. In this way, the
non-trivial interplay between local and global quantities in quantum
field theory will be checked {\it experimentally}. If the analysis
of Sect.2 is correct, the physically realized form of relativity
contains a preferred frame and one might be able to detect the
predicted non-zero density flow of energy-momentum in a moving
frame.

\section{The vacuum as a medium}

As anticipated, by assuming eq.(\ref{density}), for a moving
observer the physical vacuum looks like an ethereal medium with a
non-zero momentum density along the direction of motion. To estimate
the possible observable consequences, we shall adopt Eckart's
thermodynamical treatment \cite{eckart} of relativistic media where
the relevant quantities are the energy-momentum tensor
$W^{\alpha\beta}$ and the 4-velocity vector $u^\mu$ of the medium.

In this context, it is natural to start from a 4-velocity of the
vacuum medium ${u}^\mu(\Sigma)\equiv(1,0,0,0)$ for an observer at
rest in $\Sigma$. It is less obvious, however, to deduce its value
for a moving frame S' since the simplest choice of defining
$u^\mu(S')=\Lambda^{\mu}_{\nu}~{u}^\nu(\Sigma)$ as for an ordinary
medium, in terms of the Lorentz transformation matrix
$\Lambda^{\mu}_{\nu}$ that connects S' to $\Sigma$, can hardly be
accepted. In fact, if this were the correct transformation law, the
motion with respect to $\Sigma$ could be detected on a pure
kinematical basis regardless of the value of the vacuum energy $E_0$
that, in the quantum theory, represents the only relevant quantity
that can possibly determine a non-Lorentz invariant vacuum state.
For this reason, using this quantum input in the classical analysis,
one deduces that the motion with respect to $\Sigma$ cannot induce
any kinematical change in the description of the ethereal medium
itself as if it were seen simultaneously at rest in all frames. This
means to fix
\begin{equation} \label{all} u^\mu(S')\equiv(1,0,0,0)\end{equation}
(whatever the S' frame) so that the $u^\mu$ of the vacuum does not
transform as the 4-velocity of ordinary media and eq.(\ref{all})
should be interpreted as an external constraint on the structure of
the vacuum\cite{lorentz}. In this sense, for any moving observer S',
the vacuum medium appears at rest according to Eckart's definition
$u^\mu(S')\equiv(1,0,0,0)$ but not according to Landau's definition
since $W^{0i} \neq 0$ \cite{csernai}. The two criteria coincide only
for the observer at rest in $\Sigma$.

With such representation of the vacuum medium, by introducing the
heat flow 4-vector $q^\alpha\equiv-
s^{\alpha}_{\beta}W^{\beta\gamma}u_\gamma$, where
$s^{\alpha}_{\beta}=\delta^{\alpha}_{\beta} + u^\alpha u_\beta$, and
using eq.(\ref{all}), one finds $q^i=W^{0i}$. Therefore, from the
general relation between $q^\alpha$ and the temperature $T$
\cite{eckart}\begin{equation} \label{flow} q^\alpha= -\kappa
s^{\alpha\beta}({{\partial T}\over{\partial x^\beta}} + T u^\gamma
{{\partial u_\beta}\over{\partial x^\gamma}}) \end{equation}
($\kappa$ being the thermal conductivity of the medium), by using
again eq.(\ref{all}), one can define an effective temperature
gradient through the relation \begin{equation} \label{gradient}
{{\partial T }\over {\partial x^i}}\equiv-{{W^{0i}}\over{\kappa_0}}
 \end{equation} Here $\kappa_0$ is an unknown parameter, introduced for dimensional
reasons, that plays the role of vacuum thermal conductivity. Since
its value is unknown, the effective thermal gradient is left as an
entirely free quantity whose magnitude can be constrained by
experiments.

Formally, eq.(\ref{gradient}) is the same type of relation that one
finds in ref.\cite{eckart}. Notice, however, the basic conceptual
difference. There, one starts from a real, external temperature
gradient ${{\partial T }\over {\partial x^i}}$ to determine the heat
flow $q^i=W^{0i}$ in an ordinary medium. Here, we are starting from
the vacuum momentum density $W^{0i}=q^i$ in eq.(\ref{density}) to
define an effective ${{\partial T }\over {\partial x^i}}$. This
gradient emerges therefore as a consequence of the motion with
respect to $\Sigma$ and could induce different effects on moving
bodies. For instance, for a small temperature gradient, one expects
a pure thermal conduction in a strongly bound system, as a solid or
a liquid, and the possibility of convective currents in a loosely
bound system as a gas.

A possible objection to the previous picture is that the vacuum
should not be represented as an ordinary medium but rather as a
superfluid, see e.g. ref.\cite{volovik}. As this would carry no
entropy, moving bodies should feel no friction and thus there could
be no vacuum momentum flow and no thermal gradient. From this
perspective, the only possible condition consistent with a perfect
superfluid behaviour would be to fix $E_0=0$. On the other hand, it
is also known that in superfluid $^4$He, for any $T\neq 0$, in
addition to the pure superfluid component, there must be a small
fraction of `normal' fluid to explain the tiny residual friction
measured in the experiments. For this reason, a non-zero vacuum
energy $E_0$, that in a moving frame gives rise to the momentum
density $W^{0i}$ eq.(\ref{density}) and to the effective thermal
gradient eq.(\ref{gradient}), is equivalent to assume a small
non-superfluid component of the vacuum.

\section{The experimental test}

The effective thermal gradient eq.(\ref{gradient}), if capable to
generate convective currents in a loosely bound system as a gas,
could in principle be detected by a slight anisotropy of the speed
of light. This can be understood, in very simple terms, by
introducing the refractive index ${\cal N}$ of the gas and the time
$t$ spent by light to cover some given distance $L$. By assuming
isotropy, one finds $t={\cal N}L/c$. This can be expressed as the
sum of $t_0=L/c$ and $t_1=({\cal N}-1)L/c$ where $t_0$ is the same
time as in the vacuum and $t_1$ represents the average  time of
which light is ``slowed'' by the presence of matter. If there are
convective currents in the gas, so that $t_1$  is different in
different directions, one expects an anisotropy of the speed of
light proportional to $({\cal N}-1)$.

For this reason, one should try to design an experiment where two
orthogonal optical cavities are filled with a gas and study the
frequency shift $\Delta \nu$ between the two resonators that gives a
measure of the anisotropy of the two-way speed of light
$\bar{c}(\theta)$. As anticipated, the presence of a small
temperature gradient should give rise to two basically different
effects:

~~~~a) a pure thermal conduction in the solid parts of the
apparatus. This can affect differently the cavity length (and thus
the cavity frequency) upon active rotations of the apparatus or
under the Earth's rotation and can be preliminarily evaluated and
subtracted out by running the experiment in the vacuum mode, i.e.
when no gas is present inside the cavities. The precise experimental
limits from ref.\cite{hermann} show that any such effect can be
reduced to the level $10^{-15}-10^{-16}$.

~~~~b) small convective currents of the gas {\it inside} the
cavities that can induce a slight anisotropy
$\Delta\bar{c}_\theta\equiv \bar{c}(\pi/2)- \bar{c}(0)$ in the
two-way speed of light. On the base of the simple argument given
above, the characteristic signature would be to measure a light
anisotropy that, in two gaseous media of refractive index ${\cal N}$
and ${\cal N}'$, scales as \begin{equation} \label{scale} {{
\Delta\bar{c}_\theta( {\cal N} ) } \over{ \Delta \bar{c}_\theta(
{\cal N}') }} \sim {{ {\cal N}-1 }\over{ {\cal N}'-1 }}
\end{equation} On the other hand, for strongly bound systems, such
as when solid or liquid transparent media are filling the optical
cavities, a small temperature gradient should mainly induce a heat
conduction with no appreciable particle flow and thus with no light
anisotropy in the rest frame of the apparatus, consistently with the
classical experiments in glass and water.

This interpretation would be in agreement with the pattern observed
in the classical and modern ether-drift experiments, as illustrated
in refs.\cite{pla,cimento}. This suggests (for gaseous media {\it
only}) a relation of the type \begin{equation} \label{eq1}
{{\Delta\bar{c}_\theta( {\cal N})}\over{c}} \sim 3 ({\cal N}
-1)~{{V^2}\over{c^2}}\end{equation} where $V$ is the Earth's
velocity with respect to some preferred frame (projected in the
plane of the various interferometers). In fact, in the classical
experiments performed in air at atmospheric pressure, where ${\cal
N}\sim 1.000293$, the observed anisotropy was
${{\Delta\bar{c}_\theta}\over{c}}\lesssim 10^{-9}$ thus providing a
typical value $V/c\sim 10^{-3}$, as that associated with most cosmic
motions. Analogously, in the classical experiments performed in
helium at atmospheric pressure, where ${\cal N}\sim 1.000035$ (and
in a modern experiment with He-Ne lasers where ${\cal N}\sim
1.00004$), the observed effect was
${{\Delta\bar{c}_\theta}\over{c}}\lesssim 10^{-10}$ so that again
$V/c\sim 10^{-3}$. This means that, if there were a preferred frame,
by filling the optical resonators with gaseous media, the magnitude
of the signal might increase by 5$-$6 orders of magnitude with
respect to the limit $10^{-15}-10^{-16}$ placed by the present
ether-drift experiments in vacuum, namely from a typical $\Delta
\nu\lesssim 1$ Hz up to a $\Delta \nu\sim 100$ kHz.

\vskip 10 pt

\section{Summary and outlook}

Summarizing: in this paper we have considered two basically
different views of the vacuum. In a former approach, motivated by
the observation that (with the exception of an unbroken
supersymmetry) there is no known way to produce consistently a
vanishing vacuum energy, by using the Poincar\'e algebra of the
boost and energy-momentum operators one deduces that the physical
vacuum cannot be a Lorentz-invariant state and that, in any moving
frame, there should be a vacuum energy-momentum flow along the
direction of motion.

On the other hand, in an alternative picture where the vacuum is
only characterized by a suitable form of the expectation value of
the energy-momentum tensor (and the vacuum energy is considered a
spurious concept), one is driven to completely different
conclusions.

Since it is not so simple to decide between the two scenarios on a
pure theoretical ground, we have tried to work out the possible
phenomenological difference of the two approaches. To this end, we
have argued that the non-zero density flow of energy-momentum,
expected in a moving frame, should behave as an effective thermal
gradient. This might induce small convective currents in a loosely
bound system as a gas and produce a slight anisotropy in the speed
of light proportional to ${\cal N}-1$, ${\cal N}$ being the
refractive index of the gas. This picture is consistent with the
phenomenological pattern observed in the classical ether-drift
experiments, the only ones performed so far in gaseous systems (air
or helium at atmospheric pressure).

For this reason, we look forward to future, precise experimental
tests where optical cavities can be filled with gaseous media
(nitrogen, carbon dioxide, helium,..). In this way, one will be able
to study the beat note of the two resonators, look for modulations
of the signal that might be synchronous with the Earth's rotation
and check the trend in eq.(\ref{scale}). If a consistent non-zero
signal will be found, besides providing evidence for the existence
of a preferred frame, one will also give an experimental answer to
the non-trivial questions, concerning the interplay of global and
local quantities, mentioned in Sect.3. \vskip 20 pt
\centerline{{\boldsymbol Acknowledgments}}

 M. C. wishes to thank V. Branchina, F. Kleefeld, P. M.
Stevenson, T. Tsankov and D. Zappal\`a for useful discussions.

\vfill\eject


\begin{thebibliography} {99}
\bibitem{conformal}
Another possibility to impose the condition $E_0=0$ is to enlarge
the 10-parameter Poincar\'e group to the full 15-parameter conformal
group by including invariance under dilatations and the 4-parameter
acceleration transformations. However, again, a theory with unbroken
conformal invariance is not phenomenologically acceptable.
\end{thebibliography}
\end{document}